# Low-Latency Millimeter-Wave Communications: Traffic Dispersion or Network Densification?

Guang Yang, *Student Member, IEEE*, Ming Xiao, *Senior Member, IEEE*, and H. Vincent Poor, *Fellow, IEEE*

*Abstract*—Low latency is critical for many applications in wireless communications, e.g., vehicle-to-vehicle (V2V), multimedia, and industrial control networks. Meanwhile, for the capability of providing multi-gigabits per second (Gbps) rates, millimeter-wave (mm-wave) communication has attracted substantial research interest recently. This paper investigates two strategies to reduce the communication delay in future wireless networks: traffic dispersion and network densification. A hybrid scheme that combines these two strategies is also considered. The probabilistic delay and effective capacity are used to evaluate performance. For probabilistic delay, the violation probability of delay, i.e., the probability that the delay exceeds a given tolerance level, is characterized in terms of upper bounds, which are derived by applying stochastic network calculus theory. In addition, to characterize the maximum affordable arrival traffic for mm-wave systems, the effective capacity, i.e., the service capability with a given quality-of-service (QoS) requirement, is studied. The derived bounds on the probabilistic delay and effective capacity are validated through simulations. These numerical results show that, for a given sum power budget, traffic dispersion, network densification, and the hybrid scheme exhibit different potentials to reduce the end-to-end communication delay. For instance, traffic dispersion outperforms network densification when high sum power budget and arrival rate are given, while it could be the worst option, otherwise. Furthermore, it is revealed that, increasing the number of independent paths and/or relay density is always beneficial, while the performance gain is related to the arrival rate and sum power, jointly. Therefore, a proper transmission scheme should be selected to optimize the delay performance, according to the given conditions on arrival traffic and system service capability.

## I. INTRODUCTION

### A. Background

Wireless communications in millimeter-wave (mm-wave) bands (from around 24 GHz to 300 GHz) is a key enabler for multi-gigabits per second (Gbps) transmission [1]–[3]. In contrast to conventional wireless communications in sub-6 GHz bands, many appealing properties, including the abundant spectral resources, lower component costs, and highly directional antennas, make mm-wave communications attractive for future mobile communications standards, e.g., ECMA, IEEE 802.15.3 Task Group 3c (TG3c), IEEE 802.11ad standardization task group, and Wireess Gigabit Alliance (WiGig).

As an important metric for evaluating the quality of service (QoS), low latency plays a crucial role in the forthcoming fifth generation (5G) mobile communications [4]–[6], especially for various delay-sensitive applications, e.g., high-definition television (HDTV), intelligent transport system, vehicle-to-everything (V2X), machine-to-machine (M2M) communication, and real-time remote control. The overall delay in wireless communications consists of four components as follows [7], [8]: propagation delay (time for sending the one bit to its designated end via the physical medium), transmission delay (time for pushing the packet into the communication medium in use), processing delay (time for analyzing a packet header and making a routing decision), and queuing delay (the time that a packet spends in the buffer or queue, i.e., waiting for transmission). Normally, the overall delay for queuing system is dominantly determined by the queuing delay, while the contributions by the other types of delay are nearly negligible. Thus, for low-latency buffer-aided systems, the major task is to largely decrease the queuing delay.

In recent years, many efforts from different aspects have been devoted to low-latency mm-wave communications. In [4], several critical challenges and possible solutions for delivering end-to-end low-latency services in mm-wave cellular systems were comprehensively reviewed, from the perspectives of protocols at the medium access control (MAC) layer, congestion control, and core network architecture. By applying the Lyapunov technique for the utility-delay control, the problem of ultra-reliable and low-latency in mm-wave-enabled massive multiple-input multiple-output (MIMO) networks was studied in [9]. Regarding hybrid beamforming in mm-wave MIMO systems, a novel algorithm for achieving the ultra-low latency of mm-wave communications was proposed in [10], where the training time can be significantly reduced by progressive channel estimations. Furthermore, for systems with buffers at transceivers, the probabilistic delay for point-to-point mm-wave communications is analyzed in [11], where the delay bound is derived based on network calculus theory.

### B. Motivation

Due to unprecedented data volumes in mm-wave communications, the transceivers for many applications are commonly equipped with large-size buffers, such that the data arrivals that cannot be processed in time will be temporarily queued up in the buffer until corresponding service is provided. Hence, the low-latency problem for mm-wave communications with

This work was supported by EU Marie Curie Project, QUICK, No. 612652, Wireless@KTH Seed Project "Millimeter Wave for Ultra-Reliable Low-Latency Communications", and the U.S. National Science Foundation under Grants CNS-1702808 and ECCS-1647198.

Guang Yang and Ming Xiao are with the Department of Information Science and Engineering (ISE), KTH Royal Institute of Technology, Stockholm, Sweden (Email: {gy,mingx}@kth.se).

H. Vincent Poor is with the Department of Electrical Engineering, Princeton University, Princeton, NJ, USA (Email: poor@princeton.edu).

buffers can be interpreted as a delay problem in queuing systems, equivalently. By queuing theory, it is known that the key idea for effectively reducing the queuing delay is to keep lower service utilization. That is, the average arrival rate of data traffic should be less than the service rate of server as much as possible.

Commonly, low service utilization can be fulfilled mainly through two distinct methods: offloading arrival traffic and improving the service capability. In a wireless network, offloading arrival traffic can be realized by adopting the traffic dispersion scheme, and service enhancement can be realized by adopting the network densification scheme. Traffic dispersion stems from the application of distributed antenna systems (DASs) or distributed remote radio heads (RRHs) in mm-wave communications, and network densification is motivated from the trend of dense deployment for mm-wave networks. Roughly, the traffic dispersion scheme applies the "divide-and-conquer" principle, which enables parallel transmissions to fully exploit the spatial diversity, such that a large single queue (or large delay, equivalently) can be avoided. On the other hand, the network densification scheme departs from reducing the path loss, via shortening the separation distance between adjacent nodes, such that the end-to-end service capability can be improved.

Clearly, both the traffic dispersion and network densification schemes are promising and competitive candidates for low-latency mm-wave communications. Though there are many research contributions in low-latency communications based on above two principles, the existing literature focus on either the dispersion scheme (e.g., [12]–[17]) or multi-hop relaying scheme (e.g., [18]–[21]). It is not clear yet which scheme can provide better delay performance. For designing or implementing mm-wave networks, it is essential to explore the respective strengths of traffic dispersion and network densification, and know the their applicability and capability for realizing low-latency mm-wave networks. Moreover, a combination of traffic dispersion and network densification, termed as "hybrid scheme", is worthy of study. Intuitively, the hybrid strategy takes advantages of both traffic dispersion and network densification, and potentially can improve the delay performance under certain constraints.

### C. Objectives and Contributions

The main objective of our work is to investigate the potential of delay performance of three transmission schemes for mm-wave communications, i.e., traffic dispersion, network densification, and the hybrid scheme. Considering buffer-aided mm-wave systems, the delay performance is studied in terms of probabilistic delay and effective capacity, respectively. More precisely, to characterize the violation probability of delay, we derive corresponding probabilistic delay bounds by applying *stochastic network calculus* based on moment generating function (MGF). Furthermore, *effective capacity* is investigated to characterize the maximum asymptotic service capability. The main contributions can be summarized as follows:

- To the best of our knowledge, the comparison between traffic dispersion and network densification has not been performed previously, and thus the advantage of each scheme for delay performance is not yet clear. In this paper, we comprehensively investigate the respective strengths of above two strategies. Furthermore, we propose a generic hybrid scheme, i.e., a flexible combination of traffic dispersion and network densification, and study its potentials in reducing the communication delay. Thus, our work not only exhibits the benefits of two basic transmission strategies in different scenarios, but also explores the capability of the novel hybrid scheme in reducing the delay for certain scenarios.
- Using MGF-based stochastic network calculus, probabilistic delay bounds for traffic dispersion, network densification, and the hybrid scheme are derived, respectively. Compared to most of existing results that only consider homogeneous networks, we study the probabilistic delay heterogeneous settings for the sake of generality. Our work contributes to stochastic network calculus theory by extending the application of stochastic network calculus from homogeneous cases to heterogeneous scenarios. Though the extension to heterogeneous scenarios is not complicated, we still provide detailed derivations regarding probabilistic delay bounds via the MGF-based method, resulting in a self-contained paper for better illustration.
- Another contribution relative to MGF-based stochastic network calculus is its application in mm-wave networks. Actually, the research regarding delay analysis in wireless communications using stochastic network calculus is rather limited, although the theory has developed for decades. A remarkable achievement is the development of $(\min, \times)$-algebra [21]–[23], which was proposed to bridge the conventional stochastic network calculus and its applications in wireless scenarios. However, only Rayleigh fading channel is considered for discussion in related literature, while the investigation with respect to mm-wave fading characteristics, e.g., Nakagami-$m$ fading, still remains blank. In our research, we provide closed-form expressions for the MGF of the service process, specifically for Nakagami-$m$ fading channels. It is worth noting that the $(\min, \times)$-algebra and the $(\min, +)$-algebra are ultimately equivalent, since the Mellin transform of the exponential of a function is identical to the Laplace transform of that function. Compared to the $(\min, \times)$ version of stochastic network calculus, one appealing benefit of using the $(\min, +)$-algebra in wireless networks is that, transfers between the two domains in the $(\min, \times)$-algebra, namely the signal-to-noise ratio domain and the bit domain, can be circumvented, thereby simplifying analyses.
- It is known that, effective capacity can be used to analyze the maximum service capability in the asymptotic sense. Despite that significant progresses have been achieved in recent years, there however still remain several open issues for effective capacity, e.g., generic formulations for considered transmission schemes and analysis for networks with an arbitrary number of tandem servers. In our work, we show the maximum effective capacity in traffic dispersion with given sum power constraint, and identify the condition for achieving the optimum. Furthermore, although closed-form expressions of effective capacity for network densification and the hybrid scheme cannot be obtained (due to the convolution in $(\min,+)$-algebra), we derive lower and upper bounds to characterize their actual effective capacity. Numerical results demonstrate that the analytical lower and upper bounds are quite close to each other. Thus, the feasibility of using our derived bounds to capture the effective capacity of networks (partially or fully) with tandem architectures is validated.

- It is demonstrated that, traffic dispersion, network densification and the hybrid scheme have respective advantages, and resulting end-to-end delay performance depends on the sum power budget and the density of data traffic (e.g., average arrival rate). For instance, when the given sum power is large, traffic dispersion, the hybrid scheme, and network densification are suitable for the scenarios with heavy, medium, and light arrival traffic, respectively. However, when the given sum power is small, the corresponding strengths of above three schemes significantly change with respect to arrival traffic. These observations provide interesting insights for mm-wave network designs and implementations. That is, the transmission scheme for low-latency performance should be properly selected according to the density of arrival traffic and/or the feasible system gain.

The remainder of this paper is outlined as follows. In Sec. II, preliminaries for MGF-based stochastic network calculus are provided. In Sec. III, we give system models for traffic dispersion, network densification, and the hybrid scheme, respectively, and present MGF-based characterizations for traffic and service processes in mm-wave systems. For three low-latency schemes, corresponding probabilistic delay bounds are derived in Sec. IV, by applying MGF-based stochastic network calculus. In Sec. V, we give a closed-form expression for the effective capacity for traffic dispersion, and derive lower and upper bounds to characterize the effective capacity for both network densification and the hybrid scheme. Performance evaluations are presented in Sec. VI, where the derived results are validated, and the delay performance of three schemes is discussed. Conclusions are finally drawn in Sec. VII.

## II. PRELIMINARIES FOR NETWORK CALCULUS

In this section, for illustration purpose, we will review network calculus theory briefly and present preliminaries for MGF-based stochastic network calculus. More details for the presented fundamental results can be found in [24]–[29].

### A. Traffic and Service Process

Considering a fluid-flow, discrete-time queuing system with a buffer of infinite size, within time interval $[s,t]$, $0 \leq s \leq t$, the non-decreasing bivariate processes $A(s,t)$, $D(s,t)$ and $S(s,t)$ are defined as the cumulative arrival traffic to, departure traffic from, and service offered by server, respectively. We assume $A(s,t)$, $D(s,t)$ and $S(s,t)$ are stationary non-negative random processes, and their values are zeros whenever $s \geq t$.

Furthermore, the cumulative arrival and service processes are the sum of instantaneous realizations of each time slot within the given time interval. More exactly, let $a_i$ and $s_i$ represent the instantaneous values of arrival and service in the $i^{\text{th}}$ time slot, respectively, then $A(s,t)$ and $S(s,t)$ are given as

$$A(s,t) = \sum_{i=s}^{t-1} a_i \quad \text{and} \quad S(s,t) = \sum_{i=s}^{t-1} s_i, \quad (1)$$

for all $0 \leq s \leq t$, where time slots are normalized to 1 time unit.

For network calculus, the convolution and deconvolution in $(\min,+)$-algebra, denoted by $\otimes$ and $\oslash$, respectively, are two critical operators for deriving performance bounds of queuing systems. Their definitions with respect to the non-decreasing and strictly positive bivariate processes $X(s,t)$ and $Y(s,t)$ are respectively given as

$$(X \otimes Y)(s,t) \triangleq \inf_{s \leq \tau \leq t} \{X(s,\tau) + Y(\tau,t)\}$$

and

$$(X \oslash Y)(s,t) \triangleq \sup_{0 \leq \tau \leq s} \{X(\tau,t) - Y(\tau,s)\}.$$

With respect to cumulative arrival $A(s,t)$ and cumulative service process $S(s,t)$, according to network calculus, cumulative departure $D(s,t)$ is characterized as $D(s,t) \geq (A \otimes S)(s,t)$ [26], where the equality can be achieved if the system is linear [25]. Then, in terms of $A(s,t)$ and $D(s,t)$, the virtual delay at time $t$ is defined as $W(t) \triangleq \inf\{w \geq 0 : A(0,t) \leq D(0,t+w)\}$, which is further upper bounded by

$$W(t) \leq \inf\{w \geq 0 : (A \oslash S)(t+w,t) \leq 0\}. \quad (2)$$

Moreover, an appealing and important property of network calculus theory is the capability of dealing with tandem systems, where the equivalent end-to-end network service process can be computed as the $(\min,+)$ convolution of the individual service processes. More exactly, given $n$ concatenated servers, the end-to-end network service process $S_{\text{net}}(s,t)$ is given by

$$S_{\text{net}}(s,t) = (S_1 \otimes S_2 \otimes \cdots \otimes S_n)(s,t), \quad (3)$$

where $S_i(s,t)$ for any $1 \leq i \leq n$ denotes the service process on the $i^{\text{th}}$ hop.

### B. MGF-based Probabilistic Bounds

For queuing systems with stochastic traffic and/or service processes, the MGF-based stochastic network calculus [29] is used to effectively characterize the probabilistic delay. Among various MGF-based approaches for stochastic processes, the Chernoff bound is widely used in stochastic network calculus for deriving probabilistic bounds. More exactly, for random variable $X$ and given $x$, the Chernoff bound on $\mathbb{P}(X \geq x)$ is given as $\mathbb{P}(X \geq x) \leq e^{-\theta x}\mathbb{E}\left[e^{\theta X}\right] = e^{-\theta x}\mathbb{M}_X(\theta)$, where $\mathbb{E}[Y]$ and $\mathbb{M}_Y(\theta)$ are the mean value and the MGF (or the Laplace transform) with respect to $Y$, respectively, and $\theta$ is a positive free parameter. For any stochastic process $X(s,t), t \geq s$, the MGF of $X$ for any $\theta \geq 0$ is defined as $\mathbb{M}_X(\theta,s,t) \triangleq \mathbb{E}\left[e^{\theta X(s,t)}\right]$ [30]. Moreover, $\overline{\mathbb{M}}_X(\theta,s,t) \triangleq \mathbb{M}_X(-\theta,s,t) = \mathbb{E}\left[e^{-\theta X(s,t)}\right]$ is also defined, likewise.

The following two inequalities regarding the MGF, associated with convolution $\otimes$ and deconvolution $\oslash$, respectively, are extensively applied in MGF-based network calculus [30]. More exactly, let $X(s,t)$ and $Y(s,t)$ be independent random processes, we have

$$\overline{\mathbb{M}}_{X \otimes Y}(\theta, s, t) \leq \sum_{u=s}^{t} \overline{\mathbb{M}}_X(\theta, s, u) \cdot \overline{\mathbb{M}}_Y(\theta, u, t) \quad (4)$$

and

$$\mathbb{M}_{X \oslash Y}(\theta, s, t) \leq \sum_{u=0}^{s} \mathbb{M}_X(\theta, u, t) \cdot \overline{\mathbb{M}}_Y(\theta, u, s). \quad (5)$$

Based on (4) and (5), many properties for MGF-based stochastic network calculus are summarized in [30].

For the tandem network shown in (3), the corresponding MGF is written as $\overline{\mathbb{M}}_{S_{\text{net}}}(\theta, s, t)$ for the system with $n$ tandem servers, which is bounded by

$$\overline{\mathbb{M}}_{S_{\text{net}}}(\theta, s, t) \triangleq \overline{\mathbb{M}}_{S_1 \otimes S_2 \otimes \cdots \otimes S_n}(\theta, s, t)$$
$$\leq \sum_{s \leq u_1 \leq \cdots \leq u_{n-1} \leq t} \prod_{i=1}^{n} \overline{\mathbb{M}}_{S_i}(\theta, u_{i-1}, u_i), \quad (6)$$

where $u_0 = s$ and $u_N = t$, and $S_i, i = 1, \ldots N$ denotes the service process on each hop. (6) is obtained via applying the union bound and independence assumption [30].

Assuming independent arrival traffic and service process, in terms of MGF-based characterizations for cumulative arrival traffic and cumulative service process, i.e., $\mathbb{M}_A(\theta, s, t)$ and $\overline{\mathbb{M}}_S(\theta, s, t)$, respectively, we define $\mathsf{M}_{A,S}(\theta, s, t)$ as

$$\mathsf{M}_{A,S}(\theta, s, t) \triangleq \sum_{u=0}^{\min(s,t)} \mathbb{M}_A(\theta, u, t) \cdot \overline{\mathbb{M}}_S(\theta, u, s). \quad (7)$$

Then, based on (7), the violation probability is bounded as

$$\begin{aligned}
\mathbb{P}(W(t) \geq w) &\leq \mathbb{P}((A \oslash S)(t + w, t) \geq 0) \\
&\leq \inf_{\theta > 0} \mathbb{M}_{A \oslash S}(\theta, t + w, t) \\
&\leq \inf_{\theta > 0} \mathsf{M}_{A,S}(\theta, t + w, t) \triangleq \epsilon^w,
\end{aligned} \quad (8)$$

where the Chernoff bound and the inequality in (5) are used, and $\epsilon^w$ denotes the violation probability of delay. The last line in (8) is obtained by applying the inequality for $\mathbb{M}_{A \oslash S}$ (refer to (5)) and the definition in (7). Solving $w$, we can obtain [21], [30]

$$w = \inf \left\{ \tilde{w} \geq 0 : \inf_{\theta > 0} \{\mathsf{M}_{A,S}(\theta, t + \tilde{w}, t)\} \leq \epsilon^w \right\}. \quad (9)$$

## III. SYSTEM DESCRIPTIONS AND MGF-BASED TRAFFIC/SERVICE CHARACTERIZATIONS

### A. System Model

Throughout this paper, we assume a constant arrival rate $\rho$ for the incoming data traffic[1], i.e., $A(s,t) = \rho \cdot (t-s)$ for any $0 \leq s \leq t$. Two schemes, i.e., traffic dispersion and network densification, are illustrated in Fig. 1. They work as follows:

---
[1]Constant-rate arrival traffic is mainly considered throughout this work for simplifying analyses, while discussions on the performance associated with stochastic arrival are also attached.

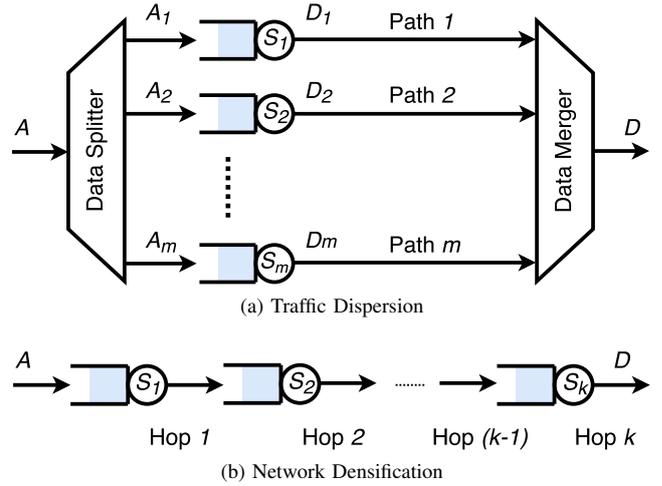

Fig. 1. Illustrations of two schemes for reducing latency for mm-wave communications: (a) traffic dispersion, and (b) network densification.

- For traffic dispersion (as shown in Fig. 1a), the original arrival traffic is firstly partitioned into multiple sub-streams by the data splitter. More precisely, given a set of deterministic splitting coefficients $(z_1, z_2, \cdots, z_n)$, where $\sum_{i=1}^{n} z_i = 1$ and $z_i \in (0,1)$ for any $1 \leq i \leq n$, the $i^{\text{th}}$ sub-stream $A_i(s,t)$ is obtained as $A_i(s,t) = z_i \cdot A(s,t)$. Then, each sub-stream gets served and delivered towards the receiver through the given path, independently. Finally, the receiver combines all sub-streams through the data merger from different paths, thereby forming the output traffic. Thanks to narrow beams (or highly directional antennas) in mm-wave communications, the inter-channel interference is negligible, and hence it is reasonable to assume independence for multiple propagation paths[2]. The principle of traffic dispersion is to decompose the original heavy arrival traffic into multiple lighter ones, thereby to avoid a large queue in the buffer.
- For network densification (as shown in Fig. 1b), multiple relay nodes[3] as servers are deployed along the source-destination transmission path. Due to the concatenation of relaying nodes, the output traffic from one relay can be treated as the input traffic for the subsequent connected relay. The application of multi-hop relaying follows trend of ultra-dense mm-wave networks. Likewise, it is feasible to assume independent channel conditions on multiple hops, thanks to high directivity and propagation properties for mm-wave communications. In contrast to the principle of traffic dispersion, the mechanism of network densification is deploying a large number of relay nodes between the given source and destination, which can reduce the path loss on each hop via shortening the separation distance between adjacent nodes, thereby increase the end-to-end service capability.

---
[2] In DAS or RRH, a large number of antenna elements can be divided into clusters and physically isolated. We also note that for the short wavelength mm-wave, lots of antenna can be co-located in a small area [3]. With the aid of proper beamforming techniques (precoding at the transmitter and signal shaping at the receiver), the orthogonality among distinct beams with high directivity is enabled, thereby providing multiple interference-free parallel paths from different clusters in mm-wave bands, where channel fading characteristics are independent due to the separation among distinct paths [31]–[33].

[3]Full-duplex relay nodes are used throughout this paper, where the self interference is ignored for simplifying analysis.

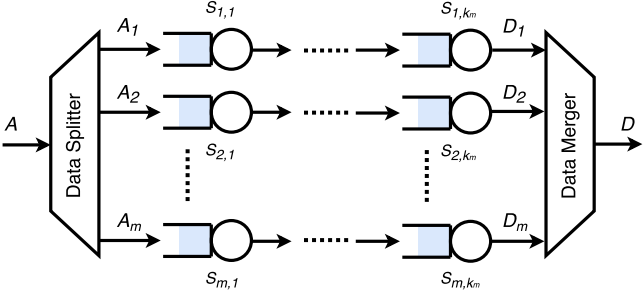

Fig. 2. Hybrid scheme for low-latency mm-wave communications.

Besides, combining the benefits of traffic dispersion and network densification, we consider hybrid scheme as shown in Fig. 2. The original arrival traffic is firstly divided into multiple sub-streams by data splitter. Subsequently, these sub-streams are allocated with independent paths for data transmission, and each path consists of multiple relay nodes. It is evident that, this combination takes advantages of traffic dispersion and network densification, i.e., offloading the arrival traffic and enhancing the service capability.

For the propagation characteristic in mm-wave bands, it is known that the small-scale fading in mm-wave channels is very weak [1], in contrast to that in sub-6 GHz bands. For the sake of tractability, the amplitude of the channel coefficient in mm-wave bands is commonly modeled as a Nakagami-$m$ random variable, as in [34], [35]. For simplicity, we assume the small-scale fading channel with Nakagami-$m$ distribution is independent and identically distributed (i.i.d.) over time in terms of blocks, i.e., i.i.d. block fading. Given separation distance $l$ and path loss exponent $\alpha$, the capacity of the mm-wave channel is given as

$$C = B \log_2 \left(1 + \xi \gamma l^{-\alpha}\right), \quad (10)$$

where $\gamma$ denotes the transmit power normalized by the background noise, $B$ is the bandwidth, and the random variable $\xi$ represents the channel gain, which follows the gamma distribution, i.e., $\xi \sim \Gamma\left(M, M^{-1}\right)$, with respect to Nakagami parameter $M$. The p.d.f. of the gamma-distributed $\xi$ is given as

$$f\left(x; M, M^{-1}\right) \triangleq \frac{x^{M-1} \exp(-Mx)}{M^{-M}\Gamma(M)},$$

where $\Gamma(z) \triangleq \int_0^\infty z^{t-1} \exp(-t)\, dt$ denotes the gamma function for $\Re(z) > 0$.

### B. MGF Bounds for Service Processes

According to (8), it can be found that, $\mathbb{M}_A(\theta, s, t)$ and $\overline{\mathbb{M}}_S(\theta, s, t)$ are required to compute the probabilistic delay bound. Regarding $\mathbb{M}_A(\theta, s, t)$, for simplifying illustration, deterministic arrivals with constant rate $\rho > 0$ are assumed in this paper, such that $\mathbb{M}_A(\theta, s, t)$ with respect to free parameter $\theta > 0$ can be written as

$$\mathbb{M}_A(\theta, s, t) = \exp(\theta \cdot \rho \cdot (t - s)) \triangleq \mu^{t-s}(\theta), \quad (11)$$

where $\mu(\theta) \triangleq \exp(\theta\rho)$. Moreover, for cumulative service process $S(s,t)$, in terms of channel capacity by (10), we have $S(s,t) = \sum_{q=s}^{t-1} C^{(q)} = \eta \sum_{q=s}^{t-1} \ln\left(1 + \xi^{(q)} \gamma l^{-\alpha}\right)$, where $\eta \triangleq B \log_2 e$, and $C^{(q)}$ denotes the instantaneous channel capacity with respect to gamma-distributed $\xi^{(q)}$ at time $q$. Then, the corresponding MGF can be written as

$$\overline{\mathbb{M}}_S(\theta, s, t) = \left(\mathbb{E}\left[\left(1 + \xi\gamma l^{-\alpha}\right)^{-\eta\theta}\right]\right)^{t-s} \triangleq \mathcal{U}_C^{t-s}(\eta\theta), \quad (12)$$

where i.i.d. Nakagami-$m$ fading across the time dimension is assumed. Here, $\mathcal{U}_C(x)$ for $x > 0$ is defined as

$$\mathcal{U}_C(x) \triangleq \left(\frac{Ml^\alpha}{\gamma}\right)^M U\left(M, 1 + M - x, \frac{Ml^\alpha}{\gamma}\right),$$

where $U(a,b,z) \triangleq \Gamma(a) \int_0^\infty \exp(-zt) t^{a-1} (1+t)^{b-a-1} dt$ denotes the *confluent hypergeometric Kummer U-function*.

Note that, with stochastic service processes, it is intractable to obtain the closed-form probability distribution function (p.d.f.) for delay. Besides, the MGF-based approach gives bounds, instead of the actual delay. Therefore, for schemes to be investigated subsequently, it is infeasible and meaningless to formulate optimization problems with certain given constraints to optimize the actual delay performance. In this sense, our work mainly aims to characterize the delay performance via bounds, rather than to optimize the schemes.

## IV. PROBABILISTIC DELAY BOUNDS FOR LOW-LATENCY TRANSMISSION SCHEMES

In this section, we mainly focus on the performance analysis of traffic dispersion and network densification, and derive upper bounds for probabilistic delay. Subsequently, based on the derived results for above two basic schemes, the delay bound for the hybrid scheme is also presented.

### A. Delay Bound for Traffic Dispersion

As shown in Fig. 1a, assuming $m \geq 1$ independent paths for the traffic dispersion scheme, associated with a set of deterministic splitting coefficients $(z_1, z_2, \cdots, z_m)$, where $\sum_{i=1}^m z_i = 1$ and $z_i \in (0,1)$ for any $1 \leq z_i \leq m$, the original cumulative arrival $A(s,t) = \rho(t-s)$ is decomposed into several sub-streams $A_i(s,t)$ for $1 \leq i \leq m$, i.e.,

$$A_i(s,t) = z_i A(s,t) = (z_i \cdot \rho) \cdot (t-s) \triangleq \rho_i(t-s).$$

In this sense, we have $\rho = \sum_{i=1}^m \rho_i$. Then, for each sub arrival traffic $A_i(s,t)$, we similarly have

$$\mathbb{M}_{A_i}(s,t) = \exp(\theta \cdot \rho_i \cdot (t-s)) \triangleq \mu_i^{t-s}(\theta). \quad (13)$$

For each transmission path, the channel capacity of the $i^{\text{th}}$ path at time $q$ is given as $C_i'^{(q)} = B \log_2\left(1 + \xi_i^{(q)} \gamma_i l^{-\alpha}\right)$, where separation distance $l$ is assumed for each path, and $\xi_i^{(q)}$ and $\gamma_i$ denote the instantaneous gamma-distributed channel gain and normalized transmit power on the $i^{\text{th}}$ path, respectively. Following (12), the MGF of $S'_i(s,t) \triangleq \sum_{q=s}^{t-1} C'^{(q)}_i$ can be written as

$$\overline{\mathbb{M}}_{S'_i}(\theta, s, t) = \mathcal{U}_{C'_i}^{t-s}(\theta) \triangleq \psi_i^{t-s}(\theta). \quad (14)$$

Without loss of generality, considering heterogeneous traffic dispersion, i.e., $\mu_i(\theta) \neq \mu_j(\theta)$ or $\psi_i(\theta) \neq \psi_j(\theta)$ may hold for $1 \leq i \neq j \leq m$, an upper bound on the violation probability for traffic dispersion is given in the following theorem.

**Theorem 1.** *Let $W(t) \triangleq \max\{W_1(t), W_2(t), \cdots, W_m(t)\}$ be the delay for the traffic dispersion scheme with $m$ independent paths, where $W_i(t)$ denotes the delay on the $i^{\text{th}}$ path. Then, for any $w \geq 0$, the probabilistic delay is bounded as follows:*

$$\mathbb{P}(W(t) \geq w) \leq 1 - \prod_{i=1}^{m}\left[1 - \inf_{\theta_i > 0}\left\{\frac{\psi_i^w(\theta_i)}{1 - \mu_i(\theta_i)\psi_i(\theta_i)}\right\}\right]^+,$$

*whenever the stability condition $\mu_i(\theta_i)\psi_i(\theta_i) < 1$ holds for some $\theta_i > 0$ and all $1 \leq i \leq m$, where $[x]^+ \triangleq \max\{x, 0\}$ for $x \in \mathbb{R}$.*

*Proof:* Please refer to Appendix A. ∎

We notice from Theorem 1 that the definition of $W(t)$ indicates the synchronization constraint. The traffic dispersion scheme discussed here actually acts as a special variant of general *fork-join* systems [16], [17], since all tasks of a job start execution at the same time, and the job is not completed until the final task leaves the system.

In Theorem 1, the stability condition, i.e., $\mu_i(\theta_i)\psi_i(\theta_i) < 1$ for all $1 \leq i \leq m$, should be satisfied to obtain the above probabilistic delay bound. This stability condition stems from the fact that, to avoid infinite delay, the arrival rate of each sub-stream cannot exceed the service capability provided on the corresponding path. Furthermore, Theorem 1 tells that, the path with higher service utilization, i.e., higher $\mu_i(\theta_i)\psi_i(\theta_i)$, is the main contributor to large delay.

With homogeneous settings, i.e., $\mu_i(\theta) = \mu(\theta)$ and $\psi_i(\theta) = \psi(\theta)$ for all $1 \leq i \leq m$, we have the following corollary.

**Corollary 1.1.** *For homogeneous traffic dispersion, given any $w \geq 0$, the probabilistic delay bound of Theorem 1 is given as*

$$\mathbb{P}(W(t) \geq w) \leq 1 - \left[\left(1 - \inf_{\theta > 0}\left\{\frac{\psi^w(\theta)}{1 - \mu(\theta)\psi(\theta)}\right\}\right)^m\right]^+,$$

*whenever $\mu(\theta)\psi(\theta) < 1$ holds for some $\theta > 0$.*

Corollary 1.1 demonstrates that, with fixed $\mu(\theta)$ and $\psi(\theta)$, the upper bound for the violation probability $\mathbb{P}(W(t) \geq w)$ grows with increasing $m$. The observation coincides with the result mentioned in [16] and [17] that the delay roughly scales up linearly with the number of independent paths, especially when the end-to-end delay on each path is small.

It is worth mentioning that Theorem 1 and Corollary 2.1 are built on the assumption that deterministic arrival is applied, such that the independence among different $W_i(t)$ is preserved. However, when considering stochastic arrival traffic, the independence of $m$ flows after splitting does not hold, and thus $\mathbb{P}(W(t) \geq w) \neq 1 - \prod_{i=1}^{m}(1 - \mathbb{P}(W_i(t) \geq w))$. To address the difficulty induced by the dependence among stochastic sub-streams, we resort to a union bound for $\mathbb{P}(W(t) \geq w)$, i.e.,

$$\mathbb{P}(W(t) \geq w) \leq \min\left\{1, \sum_{i=1}^{m}\mathbb{P}(W_i(t) \geq w)\right\}.$$

Therefore, when $\mu_i(\theta)$ for all $1 \leq i \leq m$ are not independent, the results in Theorem 1 and Corollary 2.1 should be changed respectively to

$$\mathbb{P}(W(t) \geq w) \leq \min\left\{1, \sum_{i=1}^{m}\inf_{\theta_i > 0}\left\{\frac{\psi_i^w(\theta_i)}{1 - \mu_i(\theta_i)\psi_i(\theta_i)}\right\}\right\}$$

and

$$\mathbb{P}(W(t) \geq w) \leq \min\left\{1, m\inf_{\theta > 0}\left\{\frac{\psi^w(\theta)}{1 - \mu(\theta)\psi(\theta)}\right\}\right\},$$

where the asymptotic tightness of union bounds is roughly identical to that in Theorem 1 and Corollary 2.1, since

$$1 - \prod_{i=1}^{m}(1 - \mathbb{P}(W_i(t) \geq w)) \approx \sum_{i=1}^{m}\mathbb{P}(W_i(t) \geq w) \quad (15)$$

when $\mathbb{P}(W_i(t) \geq w)$ is sufficiently small.

### B. Delay Bound for Network Densification

In the network densification scheme for multi-hop mm-wave networks, we again assume that $A(s,t) = \rho \cdot (t-s)$, such that $\mathbb{M}_A(s,t) = \mu^{t-s}(\theta)$. Moreover, for the $k$-hop relaying channel, we denote by $C''^{(q)}_i = B\log_2\left(1 + \xi_i^{(q)}\gamma_i l_i^{-\alpha}\right)$ the instantaneous channel capacity of the $i^{\text{th}}$ hop at time $q$, where $\xi_i^{(q)}$, $\gamma_i$ and $l_i$ denote the instantaneous gamma-distributed channel gain, normalized transmit power, and transmission distance on the $i^{\text{th}}$ hop, respectively. Based on (12), the MGF of $S''_i(s,t) \triangleq \sum_{q=s}^{t-1} C''^{(q)}_i$ can be written as

$$\overline{\mathbb{M}}_{S''_i}(\theta, s, t) = \mathcal{U}_{C''_i}^{t-s}(\theta) \triangleq \phi_i^{t-s}(\theta). \quad (16)$$

Considering heterogeneous multi-hop relaying, i.e., $\phi_i(\theta) \neq \phi_j(\theta)$ may hold for $1 \leq i \neq j \leq k$, we have the following theorem regarding the probabilistic end-to-end delay bound.

**Theorem 2.** *Let $W(t)$ be the end-to-end delay of a $k$-hop network. Then, for any $w \geq 0$, the probabilistic delay is bounded as follows:*

$$\mathbb{P}(W(t) \geq w)$$
$$\leq \inf_{\theta > 0}\left\{\mu^{-w}(\theta)\sum_{v=w}^{\infty}\sum_{\substack{\sum_{i=1}^{k}\pi_i = v}}\prod_{i=1}^{k}(\mu(\theta)\phi_i(\theta))^{\pi_i}\right\},$$

*whenever the stability condition $\mu(\theta)\phi_i(\theta) < 1$ holds for some $\theta > 0$ and all $1 \leq i \leq k$.*

*Proof:* Please refer to Appendix B. ∎

Similarly, the stability condition indicates that, to avoid infinite delay, the arrival rate cannot exceed the service capability provided by each hop. Besides, by Theorem 2, the hop with higher service utilization, i.e., higher $\mu(\theta)\phi_i(\theta)$, is the main contributor to a large delay.

In the following corollary, the probabilistic delay for multi-hop relaying with homogeneous settings is studied, where $\phi_i(\theta) = \phi(\theta)$ for all $1 \leq i \leq k$ are assumed.

**Corollary 2.1.** *For homogeneous network densification, given any $w \geq 0$, the probabilistic delay bound of Theorem 2 is given as*

$$\mathbb{P}(W(t) \geq w) \leq \binom{k-1+w}{w} \inf_{\theta>0} \{\phi^w(\theta)\,_2F_1(1, k+w; 1+w; \mu(\theta)\phi(\theta))\}$$

*whenever the stability condition $\mu(\theta)\phi(\theta) < 1$ holds for some $\theta > 0$. Here $_2F_1(a, b; c; z)$ is a hypergeometric function, defined as*

$$_2F_1(a, b; c; z) \triangleq \sum_{n=0}^{\infty} \frac{(a)_n (b)_n}{(c)_n} \cdot \frac{z^n}{n!},$$

*where $(x)_n$ denotes the rising Pochhammer symbol, given as*

$$(x)_n \triangleq \begin{cases} 1, & n = 0 \\ \frac{(x+n-1)!}{(x-1)!}, & n > 0 \end{cases}.$$

*Proof:* With the homogeneous setting, we obtain that

$$\mathsf{M}_{A,S''}(\theta, s, t) \leq \mu^{t-s}(\theta) \sum_{v=\tau}^{\infty} \binom{k-1+v}{v} (\mu(\theta)\phi(\theta))^v. \quad (17)$$

Regarding the infinite sum in (17), we have

$$\sum_{v=\tau}^{\infty} \binom{k-1+v}{v} (\mu(\theta)\phi(\theta))^v$$
$$= (\mu(\theta)\phi(\theta))^\tau \binom{k-1+\tau}{\tau} {}_2F_1(1, k+\tau; 1+\tau; \mu(\theta)\phi(\theta)),$$

where $\tau \triangleq \max\{s-t, 0\}$ and $_2F_1(a, b; c; z)$ is a hypergeometric function. Paired with (8) and the upper bound for $\mathsf{M}_{A,S''}(\theta, s, t)$, the theorem can be concluded by letting $s = t + w$. ∎

Corollary 2.1 demonstrates that, given fixed $\mu(\theta)$ and $\phi(\theta)$, the upper bound for the violation probability $\mathbb{P}(W(t) \geq w)$ grows when the number of hops $k$ increases, i.e., scaling as $\mathcal{O}(k^w)$. The $\mathcal{O}(\cdot)$ is defined as a set of functions $u(x)$, i.e., $\mathcal{O}(f(x)) \triangleq \{u(x) \in \mathbb{R} : \sup |u(x)/f(x)| < \infty\}$, where $f(x) \in \mathbb{R}$.

### C. Delay Bound for the Hybrid Scheme

In the hybrid scheme with $m$ ($m \geq 1$) independent transmission paths, for the arrival traffic, we assume that the sub-stream $A_i(s,t)$ given in (13). Furthermore, we denoted by $C_{i,j}^{(q)} = B \log_2\left(1 + \xi_{i,j}^{(q)} \gamma_{i,j} l_{i,j}^{-\alpha}\right)$ the instantaneous channel capacity of the $j^{\text{th}}$ hop on the $i^{\text{th}}$ transmission path at time $q$, where $\xi_{i,j}^{(q)}$, $\gamma_{i,j}$ and $l_{i,j}$ represent the instantaneous gamma-distributed channel gain, normalized transmit power and propagation distance, respectively. Similarly, according to (12), the MGF of $S_{i,j}(s,t) \triangleq \sum_{q=s}^{t-1} C_{i,j}^{(q)}$ can be written as

$$\overline{\mathbb{M}}_{S_{i,j}}(\theta, s, t) = \mathcal{U}_{C_{i,j}}^{t-s}(\theta) \triangleq \varphi_{i,j}^{t-s}(\theta). \quad (18)$$

In light of above, the probabilistic delay bound for the hybrid scheme is presented in the following theorem.

**Theorem 3.** *Assuming $m$ ($m \geq 1$) independent paths for the hybrid scheme system, with $k_i$ hops on the $i^{\text{th}}$ path for $1 \leq i \leq m$, we define $\hat{p}_i$ for any given $w \geq 0$ as*

$$\hat{p}_i \triangleq \inf_{\theta_i > 0} \left\{ \mu_i^{-w}(\theta_i) \sum_{v=w}^{\infty} \sum_{\substack{k_i \\ \sum_{j=1}^{k_i} \pi_j = v}} \prod_{j=1}^{k_i} (\mu_i(\theta_i) \varphi_{i,j}(\theta_i))^{\pi_j} \right\}.$$

*Then, the end-to-end probabilistic delay is upper bounded as*

$$\mathbb{P}(W(t) \geq w) \leq 1 - \prod_{i=1}^{m} [1 - \hat{p}_i]^+,$$

*whenever the stability condition $\mu_i(\theta_i) \varphi_{i,j}(\theta_i) < 1$ holds for some $\theta_i > 0$ and all $1 \leq i \leq m$ and $1 \leq j \leq k_i$.*

*Proof:* Applying Theorem 2 in Theorem 1 for each independent path, it is then straightforward to conclude the probabilistic delay bound for the hybrid scheme. ∎

Particularly, with homogeneous settings in the hybrid scheme, i.e., $\mu_i(\theta) = \mu(\theta)$, $\varphi_{i,j}(\theta) = \varphi(\theta)$, and $k_i = k$ for all $1 \leq i \leq m$ and $1 \leq j \leq k_i$, the result by Theorem 3 can be further reduced, which can be presented via combining Corollary 1.1 and Corollary 2.1. For brevity, results related to the homogeneous scenario are omitted.

Again, if stochastic arrival traffic is applied, the independence among sub-streams does not hold. Then upper bound in Theorem 3 should be changed to

$$\mathbb{P}(W(t) \geq w) \leq \min\left\{1, \sum_{i=1}^{m} \hat{p}_i\right\},$$

which is obtained via applying union bound (similar method in (15)).

### V. EFFECTIVE CAPACITY ANALYSIS WITH GIVEN AVERAGE SYSTEM GAIN

From the analysis of Sec. IV that, it is important to ensure that the arrival rate is below the service capability (refer to the stability conditions for three schemes). That is, the service capability characterizes the limiting potentials to deal with data traffic without causing infinite delays. Thus, in what follows, we use effective capacity [36], which is another important metric related to the delay performance departing from asymptotic service capability, to analyze three schemes.

### A. Basics for Effective Capacity

In light of the MGF of the service process by (12), the effective capacity is defined as

$$\mathcal{C}(-\theta) \triangleq \lim_{t \to \infty} \frac{\log \overline{\mathbb{M}}_S(\theta, 0, t)}{-\theta t}, \quad (19)$$

where $\theta > 0$ represents the QoS exponent, which indicates a more stringent QoS requirement for a higher $\theta$.

With certain positive $\theta$ that enables

$$\lim_{x \to \infty} \frac{\log \left( \zeta^{-1} \mathbb{P}\left( W(t) \geq x \right) \right)}{x} = -\theta \mathcal{C}\left( -\theta \right), \quad (20)$$

where $\zeta$ is the probability that the queue is not empty, the violation probability of delay, denoted by $\mathbb{P}\left( W(t) \geq w_{\max} \right)$, can be approximated as [37]–[40]

$$\mathbb{P}\left( W(t) \geq w_{\max} \right) \approx \zeta \exp\left( -\theta \mathcal{C}\left( -\theta \right) w_{\max} \right), \quad (21)$$

where $w_{\max}$ represents the maximum tolerance of delay.

In addition, for the stability consideration of first-in-first-out (FIFO) queuing systems in the asymptotic sense, according to the *Gärtner-Ellis Theorem* [41], the arrival and service process should satisfy the following condition with given $\theta$, i.e.,

$$\mathcal{R}(\theta) \triangleq \lim_{t \to \infty} \frac{\log \mathbb{M}_A(\theta, 0, t)}{\theta t} \leq \mathcal{R}^*(\theta) \triangleq \mathcal{C}(-\theta), \quad (22)$$

where $\mathcal{R}(\theta)$ in terms of QoS exponent $\theta$ is commonly termed as the "effective bandwidth" [26], and the maximum effective bandwidth, denoted by $\mathcal{R}^*(\theta)$ is characterized by the effective capacity $\mathcal{C}(-\theta)$. We notice that, the relation between effective bandwidth and effective capacity, i.e., $\mathcal{R}(\theta) \leq \mathcal{C}(-\theta)$, follows the intuition of asymptotic stability (or stability in the long-term sense), and its principle resembles the stability condition stated in Theorem 1, Theorem 2, or Theorem 3. It is worth mentioning that effective capacity $\mathcal{C}(-\theta)$ depicts the utmost service capability provided by the system, which is independent of the density of arrival traffic (refer to (19)).

According to the properties above, it is clear that, for any given QoS exponent $\theta > 0$, a larger effective capacity $\mathcal{C}(-\theta)$ not only indicates a stronger capability for serving heavier arrival traffic (see (22)), but also leads to faster decay in the probability delay (refer to (21)). Therefore, in the following analysis, we focus on the effective capacity (or the maximum effective bandwidth, equivalently) of traffic dispersion, network densification and the hybrid scheme, respectively.

In what follows, we investigate the effective capacity for traffic dispersion, network densification and the hybrid scheme. We denote by $L$ the end-to-end distance between the source and the destination (in network densification and the hybrid scheme, distance $L$ is assumed to each independent path). Besides, we assume that all transmitter are subject to a sum-power constraint $\gamma$. It is worth mentioning that the expressions of the effective capacity for networks with heterogeneous settings can be obtained by using $\overline{\mathbb{M}}_S(\theta, 0, t)$ (refer to (14) and (16) for traffic dispersion and network densification, respectively), and hence they are omitted in this section to avoid redundancy. However, we in the following analyses mainly consider homogeneous settings for three schemes, aiming at obtaining closed-form expressions for fair comparisons without loss of tractability.

### B. Effective Capacity of Traffic Dispersion

For traffic dispersion, the effective capacity of each independent path can be obtain as $\mathcal{C}_{S'_i}(-\theta) = -\frac{1}{\theta} \log \mathbb{E}\left[ \exp\left( -\theta C'_i \right) \right]$ where the independence of channel condition across the time dimension is applied. Then, considering $m$ parallel independent paths, the effective capacity of traffic dispersion is given as

$$\mathcal{C}_{S'}(-\theta) \triangleq \mathcal{C}_{\sum_{i=1}^{m} S'_i}(-\theta) = \sum_{i=1}^{m} \mathcal{C}_{S'_i}(-\theta) \\ = -\lim_{t \to \infty} \sum_{i=1}^{m} \frac{\log \overline{\mathbb{M}}_{S'_i}(\theta, 0, t)}{\theta t}. \quad (23)$$

Then, the maximum effective bandwidth can be obtained as

$$\mathcal{R}^*_{S'}(\theta) = -\frac{1}{\theta} \cdot \sum_{i=1}^{m} \log \mathbb{E}\left[ \left( 1 + \xi_i \gamma_i L^{-\alpha} \right)^{-\eta\theta} \right] \\ = -\frac{1}{\theta} \cdot \sum_{i=1}^{m} \log \left( \left( \frac{ML^\alpha}{\gamma_i} \right)^M U\left( M, 1 + M - \eta\theta, \frac{ML^\alpha}{\gamma_i} \right) \right).$$

It is worth noting that the QoS exponent $\theta$ here works for the entire system. That is, $\theta$ should make component probabilistic delays in $m$ paths all satisfy the asymptotic condition in (20). In this sense, the worst component has been considered, such that we can disregard the arrival order of data and have the overall service of traffic dispersion as a sum of individual services.

With the Nakagami-$m$ fading characteristic of the mm-wave channel, the maximum effective bandwidth $\mathcal{R}^*_{S'}(\theta)$ is given in Theorem 4.

**Theorem 4.** *Given sum power constraint $\sum_{i=1}^{m} \gamma_i = \gamma$, the maximum effective bandwidth $\mathcal{R}^*_{S'}(\theta)$ is upper bounded as*

$$\mathcal{R}^*_{S'}(\theta) \leq \frac{m \log \left( \left( \frac{MmL^\alpha}{\gamma} \right)^M U\left( M, 1 + M - \eta\theta, \frac{MmL^\alpha}{\gamma} \right) \right)}{-\theta},$$

*where the equality holds if $\gamma_i = m^{-1}\gamma$ for all $1 \leq i \leq m$.*

*Proof:* Please refer to Appendix C. ∎

Theorem 4 shows that, for the traffic dispersion scheme, the maximum effective bandwidth can be achieved when applying homogeneous settings on $m$ independent paths, i.e., $\gamma_i = m^{-1}\gamma$ for all $1 \leq i \leq m$.

### C. Effective Capacity of Network Densification

For network densification, by (3), the network service process for the multi-hop relying scheme is characterized by $S''(0,t) \triangleq (S''_1 \otimes S''_2 \otimes \cdots \otimes S''_k)(0,t)$ According the definition of effective capacity, we have

$$\mathcal{C}_{S''}(-\theta) \triangleq -\lim_{t \to \infty} \frac{\log \overline{\mathbb{M}}_{S''_1 \otimes S''_2 \otimes \cdots \otimes S''_k}(\theta, 0, t)}{\theta t}. \quad (24)$$

It is worth mentioning that, since $(\min, +)$ convolution is involved for characterizing the concatenated service in network densification, it is intractable to derive the closed-form expression for the effective capacity. Thus, we will use upper and lower bounds to characterize the effective capacity of network densification. For the sake of tractability, we in what follows consider homogeneous settings for network densification, i.e., $\gamma_i = k^{-1}\gamma$ and $l_i = k^{-1}L$ for all $1 \leq i \leq k$, and we derive the closed-form upper and lower bounds on the effective capacity.

Based on the Nakagami-$m$ fading characteristic of the mm-wave channel, the lower bound and upper bound on $\mathcal{R}^*_{S''}(\theta)$ are given in Theorem 5.

**Theorem 5.** *For homogeneous network densification with $k$ independent hops, given $\theta > 0$, the maximum effective bandwidth is upper bounded as*

$$\mathcal{R}^*_{S''}(\theta) \leq -\frac{k}{\theta} \cdot \log\left(\left(\frac{ML^\alpha}{\gamma k^{\alpha-1}}\right)^M \cdot U\left(M, 1+M-\frac{\eta\theta}{k}, \frac{ML^\alpha}{\gamma k^{\alpha-1}}\right)\right),$$

*and it is lower bounded as*

$$\mathcal{R}^*_{S''}(\theta) \geq -\frac{1}{\theta} \cdot \log\left(\left(\frac{ML^\alpha}{\gamma k^{\alpha-1}}\right)^M \cdot U\left(M, 1+M-\eta\theta, \frac{ML^\alpha}{\gamma k^{\alpha-1}}\right)\right).$$

*Proof:* Please refer to Appendix D. ∎

As we can see in Theorem 5, the upper bound meets the lower bound when $k = 1$, and the resulting maximum effective bandwidth is reduced to the closed-form expression for single-hop mm-wave networks.

### D. Effective Capacity of the Hybrid Scheme

The effective capacity of the hybrid scheme is obtained as

$$\mathcal{C}_S(-\theta) = -\lim_{t\to\infty} \sum_{i=1}^{m} \frac{\log \overline{\mathbb{M}}_{S_{i,1}\otimes\cdots\otimes S_{i,k_i}}(\theta, 0, t)}{\theta t}, \quad (25)$$

where we assume $m$ independent paths and $k_i$ relay nodes on the $i^{\text{th}}$ path ($1 \leq i \leq m$).

Again, for tractability, we consider homogeneous settings for the hybrid scheme. That is, given $m \geq 1$ independent paths and $k \geq 1$ relay nodes per path, such that $m \cdot k = n$ ($m$ or $k$ is a divisor of $n$, equivalently), we assume that $\gamma_{i,j} = n^{-1}\gamma$ and $l_{i,j} = k^{-1}L$ for all $1 \leq i \leq m$ and $1 \leq j \leq k$. Then, in the following analysis, we derive the upper and lower bounds on the effective capacity. Similarly, based on the Nakagami-$m$ fading characteristic of the mm-wave channel, the lower bound for $\mathcal{R}^*_S(\theta)$ is presented in the following theorem.

**Theorem 6.** *For the homogeneous hybrid scheme with $m$ independent paths and $k = n/m$ relay nodes per path, given $\theta > 0$, the maximum effective bandwidth is upper bounded as*

$$\mathcal{R}^*_S(\theta) \leq -\frac{n}{\theta} \cdot \log\left(\left(\frac{M(mL)^\alpha}{\gamma n^{\alpha-1}}\right)^M \cdot U\left(M, 1+M-\frac{\eta\theta}{k}, \frac{M(mL)^\alpha}{\gamma n^{\alpha-1}}\right)\right),$$

*and it is lower bounded as*

$$\mathcal{R}^*_S(\theta) \geq -\frac{m}{\theta} \cdot \log\left(\left(\frac{M(mL)^\alpha}{\gamma n^{\alpha-1}}\right)^M \cdot U\left(M, 1+M-\eta\theta, \frac{M(mL)^\alpha}{\gamma n^{\alpha-1}}\right)\right).$$

*Proof:* The theorem is immediately concluded by straightforwardly applying the variable substitution for Theorem 4 and Theorem 5, and the details are omitted for brevity. ∎

From Theorem 6, we can find that traffic dispersion and network densification can be treated as two extreme cases of the hybrid scheme, i.e., corresponding to $m = n$ and $m = 1$, respectively. When $m = n$, Theorem 6 reduces to Theorem 4.

## VI. Performance Evaluation

In this section, we provide numerical results for the probabilistic end-to-end delay bound and effective capacity discussed in Sec. III and IV, respectively. Through simulations, we firstly validate the derived bounds for probabilistic delay and effective capacity, where the respective advantages of traffic dispersion and network densification are evaluated and discussed[4]. Subsequently, for the hybrid scheme (including traffic dispersing and network densification), the factors and conditions for achieving low-latency mm-wave communications are extensively studied. For fairness considerations, the homogeneous settings presumed in Sec. IV are applied. The general system configurations are summarized as follows: the bandwidth is allocated with $B = 500$ MHz, the path loss exponent $\alpha = 2.45$, Nakagami-$m$ parameter $M = 3$, and the source-destination distance $L = 1$ km.

### A. Bound Validation

The violation probabilities of delay for traffic dispersion and network densification are illustrated in Fig. 3, where the sum transmit power is $\gamma = 85$ dB and the arrival rate is $\rho = 2$ Gbps. In Fig. 3a, for both cases $n = 2$ and $n = 4$, the probabilistic delay bound derived in Corollary 1.1 accurately characterizes the slope of the simulated result. We notice that, although the violation probability of delay decreases as the number of independent paths grows, i.e., $n$ increasing from 2 to 4, the resulting improvement is not remarkable under the given sum power $\gamma$ and arrival rate $\rho$. Likewise, in Fig. 3b, the bound by Corollary 2.1 is also able to well predict the decaying rate of violation probability. However, in contrast to the traffic dispersion scheme, increasing the relay density (equivalently, increasing the number of relays) can significantly decrease the probabilistic delay.

Fig. 4 illustrates the effective capacity for traffic dispersion and network densification, with respect to the sum power $\gamma$ varying from 50 dB to 100 dB and a given QoS exponent $\theta = 2$. Clearly, the derived lower and upper bounds of effective capacity in Theorem 5 for network densification are quite close. Thus, those bounds are capable of capturing the actual effective capacity by network densification well. We find that traffic dispersion exhibits remarkable advantages when the sum power is high (the resulting effective capacity dramatically increases with $\gamma$), while the network densification scheme adversely outperforms its counterpart when having lower sum power, e.g., $\gamma \leq 80$ dB. Furthermore, for traffic dispersion, the gain achieved by elevating $n$ becomes increasingly significant only when $\gamma$ is high. However, for network densification the gain achieved by increasing $n$ is relative steady. The findings above indicate that the benefit of traffic dispersion diminishes when the sum transmit power or the number of independent paths decreases. In this sense, the network densification is a better option for the scenarios with sparser relay deployment and lower sum power budget, and this insight is also in line with the results by comparing Fig. 3a and Fig. 3b.

---
[4] Two extreme cases, i.e., traffic dispersion and network densification, are considered only for simplifying comparison, and comprehensive results regarding the generic hybrid scheme are presented afterwards.

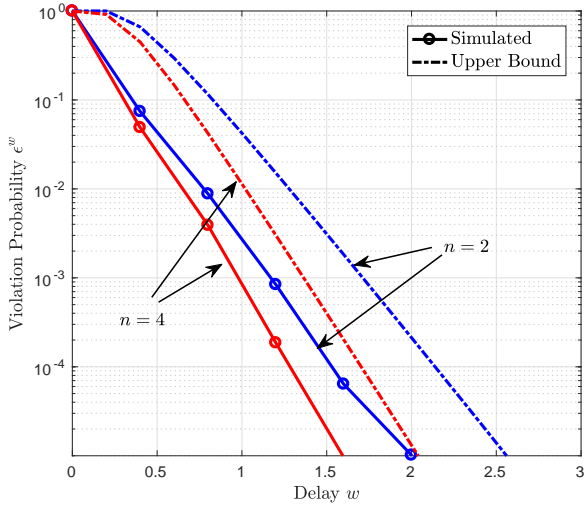

(a) traffic dispersion

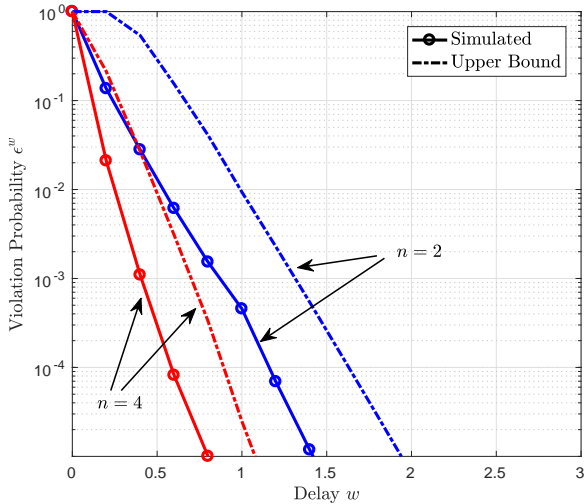

(b) network densification

Fig. 3. Violation probability $\epsilon^w$ vs. targeted delay bound $w$ for two different schemes, where $\rho = 2$ Gbps and $\gamma = 85$ dB.

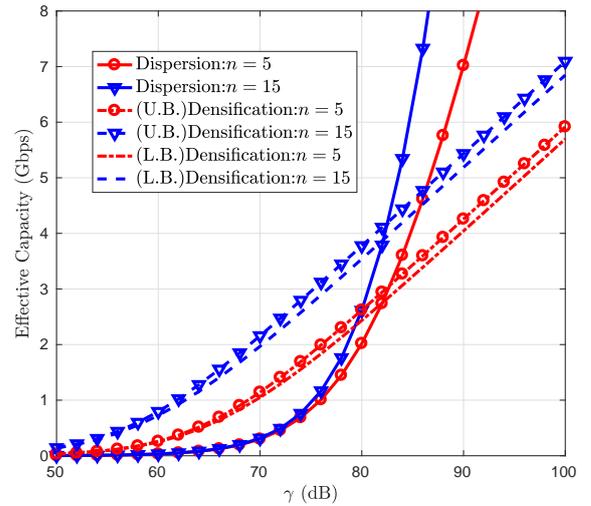

Fig. 4. Effective Capacity $\mathcal{C}(-\theta)$ vs. sum transmit power $\gamma$ for two transmission schemes, where QoS exponent $\theta = 2$. Here, U.B. and L.B. stand for "upper bound" and "lower bound", respectively.

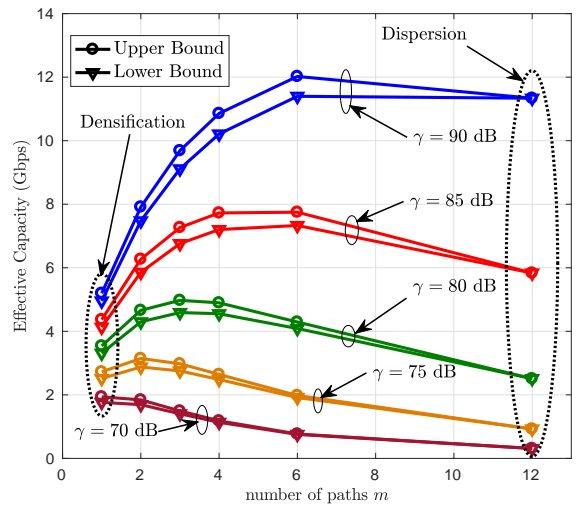

Fig. 5. Effective capacity $\mathcal{C}(-\theta)$ vs. number of independent paths $m$ for the hybrid scheme with respect to $\theta = 2$ and $n = 12$, where the number of independent paths is $m = 1, 2, 3, 4, 6$ or $12$.

### B. Simulation and Discussion

Including two extreme cases, i.e., traffic dispersion ($m = n$) and network densification ($m = 1$), the effective capacity of the hybrid scheme with $1 \leq m \leq n$ for given $n = 12$ is illustrated in Fig. 5, where the QoS exponent is $\theta = 2$, and the sum transmit power $\gamma$ varies from 70 to 90 dB. Due to the fact that the closed expression of the effective capacity for tandem networks cannot be obtained, we here again use lower and upper bounds in Theorem 6 to characterize the effective capacity of the hybrid scheme. It can be seen that, with a lower sum power, e.g., $\gamma = 70$ dB, the effective capacity decays when $m$ grows, and this observation indicates the advantage of network densification for the scenarios with a lower sum power budget. However, when $\gamma$ becomes higher, we can see that the effective capacity increases first and decreases subsequently. For instance, the maximum effective capacity is achieved at $m = 2$ in the presence of $\gamma = 75$ dB. In this case, the best solution to minimize the end-to-end delay is to assign two independent transmission paths for traffic dispersion and five relay nodes per path. It is easy to see that traffic dispersion becomes the dominant contributor to the effective capacity when $\gamma$ increases, and this tendency can be observed from the increase of the optimal number of paths. In light of the above findings, the hybrid scheme with proper configurations, i.e., the proper numbers of independent paths and hops per path, respectively, should be carefully considered to maximize the effective capacity. Also, the respective strengths of traffic dispersion and network densification revealed by Fig. 5 coincide with that from Fig. 4.

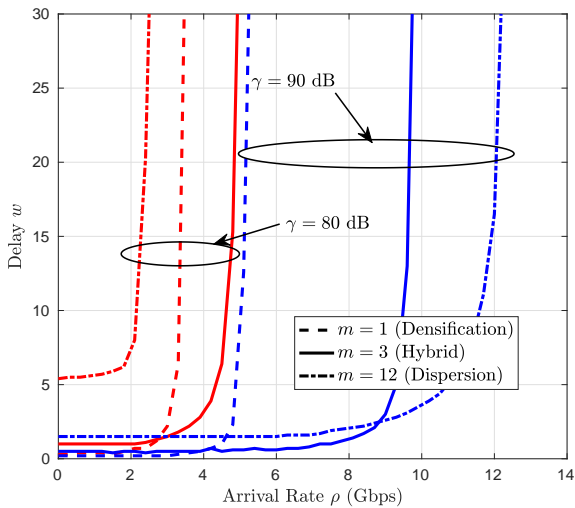

Fig. 6. Probabilistic delay $w$ vs. arrival rate $\rho$ for traffic dispersion, network densification and the hybrid scheme, respectively, with respect to violation probability $\epsilon^w = 10^{-3}$, where $n = 12$.

Coming back from effective capacity to probabilistic delay, the targeted delay tolerances versus different arrival rates for three transmission schemes i.e., traffic dispersion ($m = 12$), network densification ($m = 1$) and the hybrid scheme ($m = 3$), are provided in Fig. 6, where tolerance is given as $\epsilon^w = 10^{-3}$. For both groups in terms of different $\gamma$, clearly, the targeted delay exponentially increases when the arrival rate increases. These drastic growths result from the higher service utilization. That is, the arrival rate approaches the limiting service capability. Besides, comparing the three transmission schemes, the respective advantages demonstrated here coincide with the conclusions drawn from Fig. 5. From the perspective of probabilistic delay, we can conclude that it is critical to consider the arrival rate and the proper transmission scheme jointly to reduce the delay.

## VII. CONCLUSIONS

We have considered traffic dispersion and network densification for low-latency mm-wave communications, and have investigated their end-to-end delay performance. We have also proposed a hybrid scheme to further reduce latency in certain scenarios. Based on MGF-based stochastic network calculus and effective capacity theory, respectively, we have derived performance bounds for probabilistic delay and effective capacity for the three schemes, which have been validated through simulations. These results have demonstrated that, given the sum power budget, traffic dispersion, network densification, and the hybrid scheme show different potential in different scenarios for low-latency mm-wave communications. In addition, increasing the number of independent paths or the number of relays for network densification is always advantageous for reducing the end-to-end communication delay, while the performance gain heavily relies on the density of arrival traffic and the sum power budget, jointly. Thus, it is crucial to select the proper scheme according to the given arrival traffic and service capability.

## APPENDIX A
## PROOF OF THEOREM 1

We start with deriving the upper bound for $\mathsf{M}_{A_i, S_i'}$:

$$\mathsf{M}_{A_i, S_i'}(\theta, s, t) \overset{(a)}{\leq} \sum_{u=0}^{\min(s,t)} \mu_i^{t-u}(\theta) \psi_i^{s-u}(\theta)$$

$$\overset{(b)}{=} \mu_i^{t-s}(\theta) \sum_{v=\tau}^{s} (\mu_i(\theta) \psi_i(\theta))^v$$

$$\overset{(c)}{\leq} \mu_i^{t-s}(\theta) \sum_{v=\tau}^{\infty} (\mu_i(\theta) \psi_i(\theta))^v$$

$$= \frac{\mu_i^{t-s}(\theta) (\mu_i(\theta) \psi_i(\theta))^\tau}{1 - \mu_i(\theta) \psi_i(\theta)}, \quad (26)$$

where $\tau \triangleq \max\{s - t, 0\}$. Here, inequality $(a)$ is obtained by plugging (13) and (14) into (7). By performing the change of variable, i.e., $v = s - u$, equality $(b)$ is achieved. In $(c)$, we let $s$ go to infinity. In the final step, the geometric sum converges only when $\mu_i(\theta) \psi_i(\theta) < 1$ holds for certain $\theta$.

By the definition of $W(t)$, it is easy to obtain that

$$\mathbb{P}(W(t) \geq w) \triangleq \mathbb{P}\left(\max_{1 \leq i \leq m} \{W_i(t)\} \geq w\right)$$

$$\overset{(d)}{=} 1 - \prod_{i=1}^{m} (1 - \mathbb{P}(W_i(t) \geq w))$$

$$\overset{(e)}{\leq} 1 - \prod_{i=1}^{m} \left(1 - \inf_{\theta_i > 0} \mathsf{M}_{A_i, S_i'}(\theta_i, t + w, t)\right)$$

$$\overset{(f)}{\leq} 1 - \prod_{i=1}^{m} \left(1 - \inf_{\theta_i > 0} \left\{\frac{\psi_i^w(\theta_i)}{1 - \mu_i(\theta_i) \psi_i(\theta_i)}\right\}\right),$$

where the independence assumption among distinct paths is used to derive $(d)$, and inequality $(e)$ applies the (8), and $(f)$ follows from (26).

In addition, we notice that $\mathbb{P}(W(t) \geq w) \leq 1$ holds for any $w \geq 0$, then the theorem is concluded.

## APPENDIX B
## PROOF OF THEOREM 2

Applying (16) in (6), the MGF of $k$-hop network service process can be characterized as

$$\overline{\mathsf{M}}_{S''}(\theta, s, t) \triangleq \overline{\mathsf{M}}_{S_1'' \otimes \cdots \otimes S_k''}(\theta, s, t) \leq \sum_{\sum_{i=1}^{k} \pi_i = t-s} \prod_{i=1}^{k} \phi_i^{\pi_i}(\theta).$$

Then, it is easy to obtain that

$$\begin{aligned}
\mathsf{M}_{A,S''}(\theta, s, t) &\leq \sum_{u=0}^{\min(s,t)} \mu^{t-s}(\theta) \sum_{\sum_{i=1}^{k} \pi_i = s-u} \prod_{i=1}^{k} \phi_i^{\pi_i}(\theta) \\
&= \mu^{t-s}(\theta) \sum_{v=\tau}^{s} \sum_{\sum_{i=1}^{k} \pi_i = v} \prod_{i=1}^{k} (\mu(\theta) \phi_i(\theta))^{\pi_i} \\
&\leq \mu^{t-s}(\theta) \sum_{v=\tau}^{\infty} \sum_{\sum_{i=1}^{k} \pi_i = v} \prod_{i=1}^{k} (\mu(\theta) \phi_i(\theta))^{\pi_i},
\end{aligned} \quad (27)$$

where $\tau \triangleq \max\{s-t, 0\}$.

Similar to Theorem 1, we notice that the stability condition, i.e., $\mu(\theta)\phi_i(\theta) < 1$ for all $1 \leq i \leq k$, needs to be satisfied to guarantee the convergence of (27). Regarding the convergence, it is easy to show that, once the stability condition is met, we have $\mu(\theta)\hat{\phi}(\theta) < 1$ equivalently, where $\hat{\phi}(\theta) \triangleq \max_{1 \leq i \leq k}\{\phi_i(\theta)\}$. In this case, we obtain that

$$\sum_{\sum_{i=1}^{k} \pi_i = v} \prod_{i=1}^{k} (\mu(\theta)\phi_i(\theta))^{\pi_i} \leq \sum_{\sum_{i=1}^{k} \pi_i = v} \left(\mu(\theta)\hat{\phi}(\theta)\right)^{\sum_{i=1}^{k}\pi_i}.$$

According to combinatorics properties, we notice that

$$\begin{aligned}
&\sum_{v=\tau}^{\infty} \sum_{\sum_{i=1}^{k} \pi_i = v} \left(\mu(\theta)\hat{\phi}(\theta)\right)^{\sum_{i=1}^{k}\pi_i} \\
&= \sum_{v=\tau}^{\infty} \left(\mu(\theta)\hat{\phi}(\theta)\right)^{v} \sum_{\sum_{i=1}^{k}\pi_i = v} 1 \\
&= \sum_{v=\tau}^{\infty} \binom{k-1+v}{v} \left(\mu(\theta)\hat{\phi}(\theta)\right)^{v} \\
&\leq \sum_{v=0}^{\infty} \binom{k-1+v}{v} \left(\mu(\theta)\hat{\phi}(\theta)\right)^{v} = \left(1 - \mu(\theta)\hat{\phi}(\theta)\right)^{-k}.
\end{aligned}$$

Therefore, we can demonstrate that, $\mathsf{M}_{A,S''}(\theta, s, t)$ is convergent if the stability condition holds, since

$$\mathsf{M}_{A,S''}(\theta, s, t) \leq e^{\theta\sigma(\theta)} \mu^{t-s}(\theta) \left(1 - \mu(\theta)\hat{\phi}(\theta)\right)^{-k} < \infty.$$

Finally, with respect to (8), the theorem then can be concluded by letting $s = t + w$.

## APPENDIX C
## PROOF OF THEOREM 4

We define the function $y(r) \triangleq \log \mathbb{E}\left[(1+rX)^{-z}\right]$ for $r > 0$, where $X$ represents a positive random variable and $z$ is any positive number. Then, the first derivative of $y(r)$ with respect to $r$, written by $y'(r)$, is obtained as

$$y'(r) = -\frac{z\mathbb{E}\left[(1+rX)^{-(z+1)} X\right]}{\mathbb{E}\left[(1+rX)^{-z}\right]} < 0.$$

Besides, the second derivative of $y(r)$ with respect to $r$, written by $y''(r)$, is obtained as

$$\begin{aligned}
y''(r) &= \frac{1}{\mathbb{E}^2\left[(1+rX)^{-z}\right]} \left(z\mathbb{E}\left[\frac{X^2}{(1+rX)^{z+2}}\right] \mathbb{E}\left[\frac{1}{(1+rX)^z}\right]\right. \\
&\quad + z^2 \left(\mathbb{E}\left[\frac{X^2}{(1+rX)^{z+2}}\right] \mathbb{E}\left[\frac{1}{(1+rX)^z}\right]\right. \\
&\quad \left.\left. - \mathbb{E}^2\left[\frac{X}{(1+rX)^{z+1}}\right]\right)\right) \\
&= \frac{1}{\mathbb{E}^2\left[(1+rX)^{-z}\right]} \left(z\mathbb{E}\left[\frac{X^2}{(1+rX)^{z+2}}\right] \mathbb{E}\left[\frac{1}{(1+rX)^z}\right]\right. \\
&\quad + z^2 \left(\mathbb{E}\left[\frac{X^2}{(1+rX)^{z+2}}\right] \mathbb{E}\left[\frac{1}{(1+rX)^z}\right]\right. \\
&\quad \left.\left. - \mathbb{E}^2\left[\frac{X}{(1+rX)^{\frac{z}{2}+1}} \cdot \frac{1}{(1+rX)^{\frac{z}{2}}}\right]\right)\right) \\
&\geq \frac{z\mathbb{E}\left[(1+rX)^{-(z+2)} X^2\right] \mathbb{E}\left[(1+rX)^{-z}\right]}{\mathbb{E}^2\left[(1+rX)^{-z}\right]} > 0,
\end{aligned}$$

where the last line is obtained by applying the *Cauchy–Schwarz inequality*, i.e., $\mathbb{E}^2[AB] \leq \mathbb{E}[A^2]\mathbb{E}[B^2]$ for random variables $A$ and $B$. Since $y'(r) < 0$ and $y''(r) > 0$, it is shown that $y(r)$ is a monotonically decreasing and strictly convex function with respect to $r > 0$.

Applying Jensen's inequality, we immediately have

$$\sum_{i=1}^{m} \log \mathbb{E}\left[\left(1 + \frac{\gamma_i \xi}{L^\alpha}\right)^{-\eta\theta}\right] \geq m \log \mathbb{E}\left[\left(1 + \sum_{i=1}^{m} \frac{\gamma_i \xi}{L^\alpha}\right)^{-\eta\theta}\right],$$

where the equality is achieved if and only if $\gamma_i = \gamma_j \triangleq m^{-1}\gamma$ holds for all $1 \leq i, j \leq m$. Thus, we can easily obtain, following the same lines as above, that

$$\mathcal{R}_{S'}^*(\theta) \leq -\frac{m}{\theta} \log \mathbb{E}\left[\left(1 + \gamma\xi L^{-\alpha}\right)^{-\eta\theta}\right],$$

and the proof is completed by applying (12).

## APPENDIX D
## PROOF OF THEOREM 5

For the upper bound, by the definition of $S''(0, t)$, we have

$$S''(0, t) \leq \min_{1 \leq i \leq k}\{S_i''(0, t)\} \leq \frac{1}{k}\sum_{i=1}^{k} S_i''(0, t).$$

Then, for $\overline{\mathbb{M}}_{S''}(\theta, 0, t)$, we have

$$\overline{\mathbb{M}}_{S''}(\theta, 0, t) \geq \mathbb{E}\left[\exp\left(-\frac{\theta}{k}\sum_{i=1}^{k} S_i''(0, t)\right)\right]$$

$$= \mathbb{E}\left[\prod_{i=1}^{k} \exp\left(-\frac{\theta}{k} S_i''(0, t)\right)\right]$$

$$= \left(\mathbb{E}\left[\exp\left(-\frac{\theta}{k}\sum_{q=0}^{t-1} C_i''^{(q)}\right)\right]\right)^k$$

$$= \left(\mathbb{E}\left[\left(1 + \xi_i \left(\frac{\gamma}{k}\right)\left(\frac{L}{k}\right)^{-\alpha}\right)^{-\frac{\eta\theta}{k}}\right]\right)^{kt}$$

$$= \left(\left(\frac{ML^\alpha}{\gamma n^{\alpha-1}}\right)^M U\left(M, 1+M-\frac{\eta\theta}{k}, \frac{ML^\alpha}{\gamma k^{\alpha-1}}\right)\right)^{kt},$$

Finally, the upper bound on $\mathcal{R}_{S''}^*(\theta)$ can be obtained as

$$\mathcal{R}_{S''}^*(\theta) \leq -\frac{k}{\theta} \cdot \log\left(\left(\frac{ML^\alpha}{\gamma k^{\alpha-1}}\right)^M\right.$$
$$\left. \cdot U\left(M, 1+M-\frac{\eta\theta}{k}, \frac{ML^\alpha}{\gamma k^{\alpha-1}}\right)\right).$$

For the lower bound, based on (24) and the homogeneous settings, we have the upper bound on $\overline{\mathbb{M}}_{S''}(\theta, 0, t)$ as

$$\overline{\mathbb{M}}_{S''}(\theta, 0, t) \leq \sum_{\sum_{i=1}^k \pi_i = t} \prod_{i=1}^{k} \left(\mathbb{E}\left[\left(1+\xi_i \gamma_i l_i^{-\alpha}\right)^{-\eta\theta}\right]\right)^{\pi_i}$$

$$= \left(\mathbb{E}\left[\left(1+\xi k^{\alpha-1}\gamma L^{-\alpha}\right)^{-\eta\theta}\right]\right)^t \cdot \sum_{\sum_{i=1}^k \pi_i = t} 1$$

$$= \binom{t+k-1}{k-1}\left(\mathbb{E}\left[\left(1+\xi k^{\alpha-1}\gamma L^{-\alpha}\right)^{-\eta\theta}\right]\right)^t,$$

where the second line is achieved due to the uniformly allocated transmit power (normalized) and the identical length of each hop. Therefore, the lower bound for $\mathcal{R}_{S''}^*(\theta)$ can be obtained as

$$\mathcal{R}_{S''}^*(\theta) \geq \lim_{t\to\infty} \frac{\log\left(\binom{t+k-1}{k-1}\left(\mathbb{E}\left[\left(1+\xi k^{\alpha-1}\gamma L^{-\alpha}\right)^{-\eta\theta}\right]\right)^t\right)}{-\theta t}$$

$$\geq \frac{\log \mathbb{E}\left[\left(1+\xi k^{\alpha-1}\gamma L^{-\alpha}\right)^{-\eta\theta}\right]}{-\theta} - \lim_{t\to\infty} \frac{\log\frac{(t+k-1)^{k-1}}{(k-1)!}}{\theta t}$$

$$= \frac{\log \mathbb{E}\left[\left(1+\xi k^{\alpha-1}\gamma L^{-\alpha}\right)^{-\eta\theta}\right]}{-\theta},$$

where the inequality in the second line uses the property that

$$\binom{N}{k} \leq \frac{N^k}{k!},$$

for any integers $N \geq k \geq 0$.

Therefore, the maximum effective bandwidth for the network densification scheme is lower bounded by

$$\mathcal{R}_{S''}^*(\theta) \geq -\frac{1}{\theta}\log \mathbb{E}\left[\left(1+\xi k^{\alpha-1}\gamma L^{-\alpha}\right)^{-\eta\theta}\right]$$

$$= -\frac{1}{\theta} \cdot \log\left(\left(\frac{ML^\alpha}{\gamma k^{\alpha-1}}\right)^M\right.$$
$$\left. \cdot U\left(M, 1+M-\eta\theta, \frac{ML^\alpha}{\gamma k^{\alpha-1}}\right)\right).$$

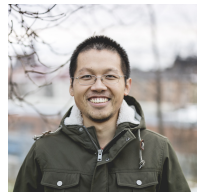

**Guang Yang** received his B.E degree in Communication Engineering from University of Electronic Science and Technology of China (UESTC), Chengdu, China in 2010, and from 2010 to 2012 he participated in the joint Master-PhD program in National Key Laboratory of Science and Technology on Communications at UESTC. He joined the Department of Information Science and Engineering at the School of Electrical Engineering and Computer Science, the Royal Institute of Technology (KTH), Stockholm, Sweden, as a Ph.D. student since September of 2013.

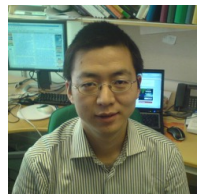

**Ming Xiao** (S'2002-M'2007-SM'2012) received Bachelor and Master degrees in Engineering from the University of Electronic Science and Technology of China, ChengDu in 1997 and 2002, respectively. He received Ph.D degree from Chalmers University of technology, Sweden in November 2007. From 1997 to 1999, he worked as a network and software engineer in ChinaTelecom. From 2000 to 2002, he also held a position in the SiChuan communications administration. From November 2007 to now, he has been in Communication Theory, school of electrical engineering, Royal Institute of Technology, Sweden, where he is currently an Associate Professor in Communications Theory. He received the best paper Awards in "IC-WCSP" (International Conference on Wireless Communications and Signal Processing) in 2010 and "IEEE ICCCN" (International Conference on Computer Communication Networks) in 2011. Dr. Xiao received "Chinese Government Award for Outstanding Self-Financed Students Studying Abroad" in March, 2007. He got "Hans Werthen Grant" from royal Swedish academy of engineering science (IVA) in March 2006. He received "Ericsson Research Funding" from Ericsson in 2010. Since 2012, he has been an Associate Editor for IEEE Transactions on Communications, IEEE Communications Letters (Senior Editor Since Jan. 2015) and IEEE Wireless Communications Letters.

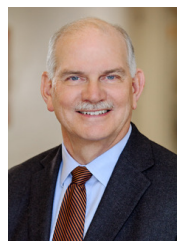

**H. Vincent Poor** (S72, M77, SM82, F87) received the Ph.D. degree in EECS from Princeton University in 1977. From 1977 until 1990, he was on the faculty of the University of Illinois at Urbana-Champaign. Since 1990 he has been on the faculty at Princeton, where he is currently the Michael Henry Strater University Professor of Electrical Engineering. During 2006 to 2016, he served as Dean of Princetons School of Engineering and Applied Science. He has also held visiting appointments at several other universities, including most recently at Berkeley and Cambridge. His research interests are in the areas of information theory and signal processing, and their applications in wireless networks, energy systems and related fields. Among his publications in these areas is the recent book *Information Theoretic Security and Privacy of Information Systems* (Cambridge University Press, 2017).

Dr. Poor is a member of the National Academy of Engineering and the National Academy of Sciences, and is a foreign member of the Chinese Academy of Sciences, the Royal Society, and other national and international academies. He received the Marconi and Armstrong Awards of the IEEE Communications Society in 2007 and 2009, respectively. Recent recognition of his work includes the 2017 IEEE Alexander Graham Bell Medal, Honorary Professorships at Peking University and Tsinghua University, both conferred in 2017, and a D.Sc. *honoris causa* from Syracuse University also awarded in 2017.